\documentclass[11pt,twoside]{article}

\usepackage{asp2006}

\usepackage{epsf}

\usepackage{lscape}

\begin{document}

\title{The VIRUS-P Exploration of Nearby Galaxies (VENGA): Survey Design and First Results}  
\author{Guillermo A. Blanc$^1$, Karl Gebhardt$^1$, Amanda Heiderman$^1$, Neal J. Evans II$^1$, Shardha Jogee$^1$, Remco van den Bosch$^1$, Irina Marinova$^1$, Tim Weinzirl$^1$, Peter Yoachim$^1$, Niv Drory$^2$, Maximilian Fabricius$^2$, David Fisher$^1$, Lei Hao$^3$, Phillip J. MacQueen$^{4}$, Juntai Shen$^3$, Gary J. Hill$^{4}$, John Kormendy$^1$}
\affil{$^1$  Department of Astronomy, The University of Texas at
  Austin,\\ 1 University Station C1400, Austin, TX 78712, USA;\\
$^2$ Max-Planck-Institut f$\ddot{u}$r extraterrestrische Physik,
  Giessenbachstra\ss e, 85748 Garching, Germany;\\ 
$^3$ Shanghai Astronomical Observatory, 80 Nandan Road, Shanghai, Shanghai 200030, China;\\
$^4$ McDonald Observatory,The University of Texas at
  Austin,\\ 1 University Station C1400, Austin, TX 78712, USA;\\ 
 \texttt{gblancm@astro.as.utexas.edu}}


\begin{abstract} 

VENGA is a large-scale extragalactic IFU survey, which maps the bulges,
bars and large parts of the outer disks of 32 nearby normal spiral galaxies. The targets are chosen
to span a wide range in Hubble types, star formation activities, morphologies, and inclinations,
at the same time of having vast available multi-wavelength coverage from the far-UV to the mid-
IR, and available CO and 21cm mapping. The VENGA dataset will provide 2D maps of the SFR,
stellar and gas kinematics, chemical abundances, ISM density and ionization states, dust extinction
and stellar populations for these 32 galaxies. The uniqueness of the VIRUS-P large field of view
permits these large-scale mappings to be performed. VENGA will allow us to correlate all these
important quantities throughout the different environments present in galactic disks, allowing
the conduction of a large number of studies in star formation, structure assembly, galactic feedback
and ISM in galaxies.

\end{abstract}

\section{Introduction}

\begin{figure}[!ht]
\begin{center}
\plotone{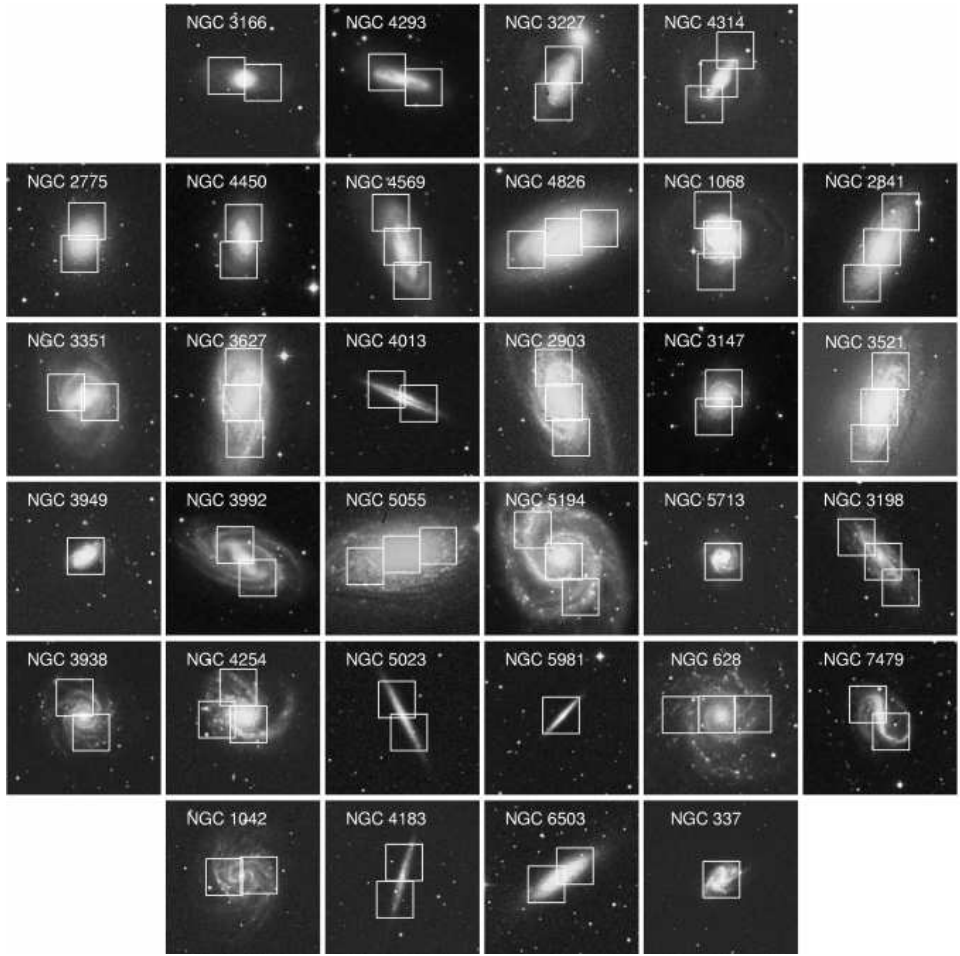}
\end{center}
\caption{DSS cutouts showing the VENGA sample of 32 nearby spirals
  ordered by RC3 morphological type from early (top left) to late
  (bottom left) types. White boxes show the VIRUS-P $1.7'\times 1.7'$
  FOV.}
\label{fig-1}
\end{figure}

In $\Lambda$CDM cosmology, the formation and evolution of galaxies
takes place in gravitational potential wells in the dark matter
distribution (DM halos). Gas accretion and merging processes
ultimately trigger star formation giving rise to galaxies
\citep{blumenthal84}. Although consensus has been reached concerning
this picture, the baryonic physics behind galaxy formation in the
centers of DM halos are aggressively debated. The triggering of star
formation and the variables that set the star formation rate (SFR)
\citep{kennicutt98b, leroy08, krumholz09, tan09}, the contribution
from different types of feedback processes (AGN, SN, stellar
radiation) at regulating the gaseous budget, structure and kinematics
of the ISM \citep{kauffmann99, croton06, thompson09}, and the role
that major and minor mergers as well as secular evolution processes
play at shaping galaxies \citep{kormendy04,weinzirl09}, are the main
current areas of research. All these processes play a major role in
determining how galaxies evolve throughout cosmic time, building up
their stellar mass and shaping their present day structure.

The detailed manner in which the above physical phenomena (star
formation, feedback, interactions, and secular evolution) proceed
ultimately determines the morphology, kinematics, stellar populations,
chemical structure, ISM structure, and star formation history (SFH) of
a galaxy. We can put constraints on these processes by obtaining
spatially resolved measurements of quantities like the SFR, stellar
and gas kinematics, stellar populations, chemical abundances (both gas
phase and photospheric), atomic and molecular gas surface densities,
etc. and studying the correlations between them. Wide field optical
integral field spectroscopy allows the measurement of many of these
quantities in nearby galaxies. IFU maps combined with multi-wavelength
broad band photometry and sub-mm and radio maps of the same galaxies
are powerful datasets to study galaxy evolution.

VENGA is a large-scale IFU survey of 32 nearby spiral galaxies (Figure
\ref{fig-1}) using the VIRUS-P spectrograph \citep{hill08, blanc09} on
the 2.7m Harlan J. Smith telescope at McDonald Observatory. The sample
spans a wide range in Hubble types, SFR, and morphologies, including
galaxies with classical and pseudo-bulges, as well as barred and
unbarred objects. Ancillary multi-wavelength data including HST
optical and NIR, Spitzer IRAC and MIPS, and far-UV GALEX imaging, as
well as mid-IR IRS spectroscopy, CO maps and HI 21cm maps, are
available for most of the sample. VENGA will allow a large number of
researchers to conduct an extensive set of studies on star-formation,
structure assembly, stellar populations, gas and stellar dynamics,
chemical evolution, ISM structure, and galactic feedback. VENGA will
also provide the best local universe control sample for IFU studies of
high-z galaxies \citep[e.g. ][]{forster09}. 

\section{Survey Design}

VIRUS-P is currently the largest field-of-view (FOV) IFU in the
world. It has 246 optical fibers (each 4.3'' in diameter) which sample
a $1.7'\times 1.7'$ field with a 1/3 filling factor. Three dithers
provide contiguous coverage of the FOV. Hence, for each VIRUS-P
pointing we obtain spectra of 738 independent spatial resolution
elements. Depending on the angular size of the targets we observe 1 to
3 poinitngs on each galaxy (see Figure \ref{fig-1}) providing full
coverage of the central parts of the galaxies and a typical sampling
of the outer disks out to 2.5 $R_{\rm{eff}}$. The spectra is obtained in
both a blue and a red setup covering the 3600\AA-5800\AA\ and
4600\AA-6800\AA\ ranges respectively, and has a spectral resolution of
5\AA\ FWHM ($\sigma_{\rm{inst}}\sim120$km s$^{-1}$). On total the
survey is composed of 74 VIRUS-P pointings over the 32 galaxies,
providing spectra for $\sim55000$ independent regions across the disks
of our targets.

VENGA also includes a secondary observing campaign using the upcoming
wide field high resolution integral field spectrograph VIRUS-W
\citep{fabricius08}, which will be conducted after the instrument is
commissioned. VIRUS-W will provide spatially resolved spectroscopy of
the VENGA sample with a spectral resolution of
$\sigma_{\rm{inst}}\sim25$km s$^{-1}$, allowing the measurement of
stellar velocity dispersions throughout the bulges, bars, and disks of
the galaxies.

\begin{figure}[!ht]
\begin{center}
\plotone{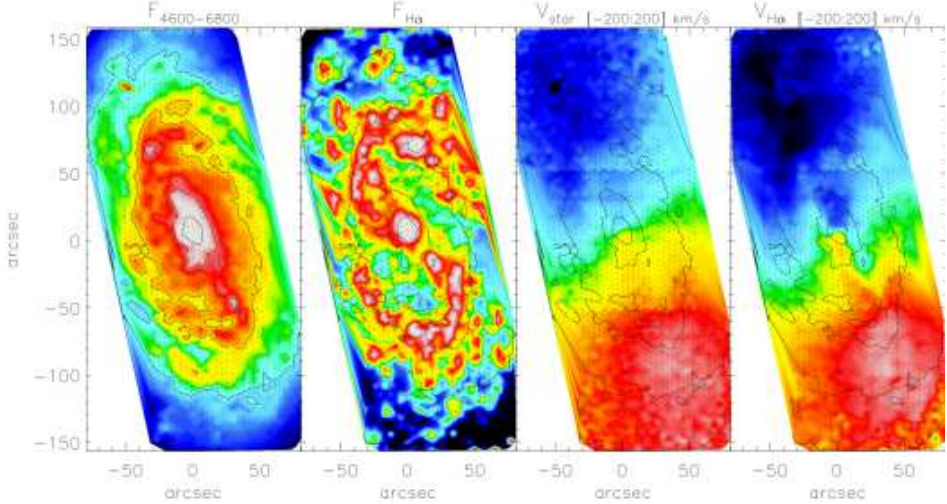}
\end{center}
\caption{VENGA maps of NGC2903. From left to right: a) Integrated
  stellar flux in the 4600\AA-6800\AA\ range. b) Flux of the H$\alpha$
  line. c) Stellar velocity. d) Ionized gas velocity from
  H$\alpha$.}
\label{fig-2}
\end{figure}

The VIRUS-P data reduction is being conducted using our custom
pipeline VACCINE, and spectral analysis (measurement of stellar
kinematics, gas kinematics, and fitting of emission lines) is being
done using a modified version of the GANDALF software \citep{sarzi06},
which includes an implementation of the Penalized Pixel-Fitting method
\citep[pPXF][]{cappellari04}. As an example of our data products
Figure \ref{fig-2} presents four VENGA maps of NGC2903 showing the
integrated stellar continuum in our red setup wavelength range, the
observed H$\alpha$ line flux, the stellar velocity, and the ionized
gas velocity as measured from the centroid of the H$\alpha$ line.

\section{First Results: The Spatially Resolved Star Formation Law}

Using the VENGA data on the central pointing of NGC5194 we investigate
the relation between the star formation rate surface density
($\Sigma_{SFR}$) and the mass surface density of gas ($\Sigma_{HI+H_2}$),
usually known as the Star Formation Law \citep[SFL, a.k.a. Schmidt Law
  or Schmidt-Kennicutt Law,][]{schmidt59, kennicutt98b}. The method
and results are reported in detail in \cite{blanc09}. 

\begin{figure}[!ht]
\begin{center}
\plotone{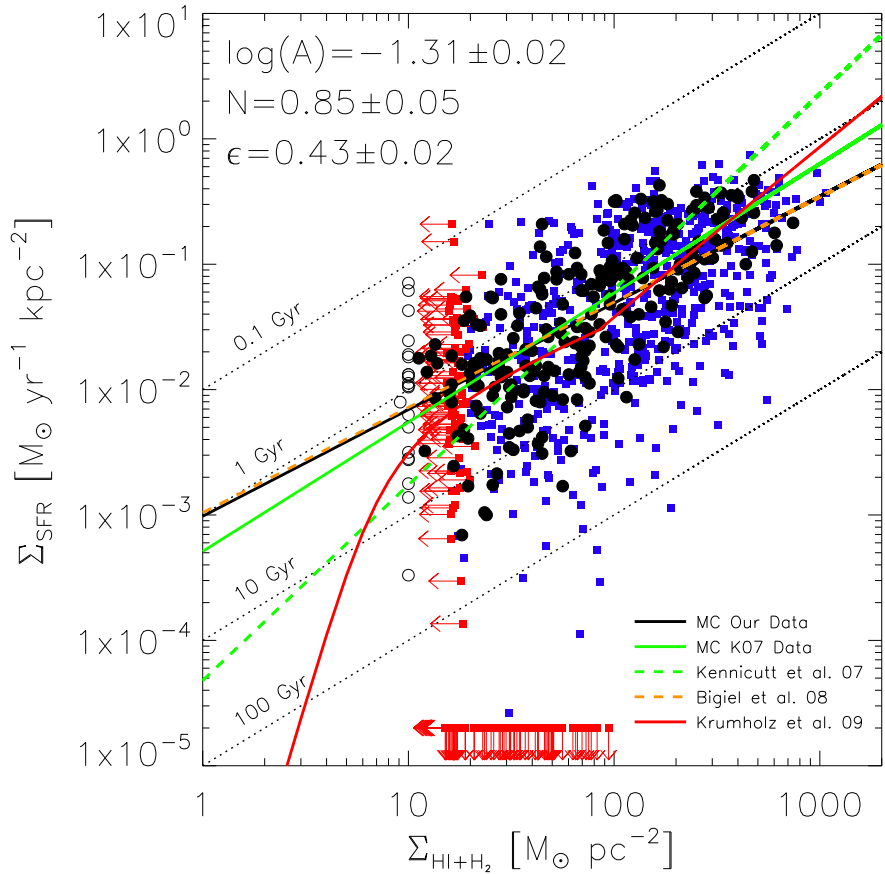}
\end{center}
\caption{Total gas SFL in the central pointing of NGC5194. The VENGA
  data ({\it blue and red squares}) shows an excellent agreement with the data
  of \cite{kennicutt07} ({\it black circles}). Lines of constant
  depletion time are shown ({\it black dotted lines}). Parameters for
  the Monte Carlo best-fit SFL to our data ({\it black solid line}) are
  reported. The Monte Carlo method best-fit to the
  \cite{kennicutt07} data is presented ({\it green solid line}) and
  has a slope of $N=1.03\pm0.08$. Also shown is the theoretical model of \cite{krumholz09}
  ({\it red solid line}).}\label{fig-3}
\end{figure}

From the VIRUS-P spectra we measured H$\alpha$, H$\beta$,
[NII]$\lambda\lambda$6548,6584, and [SII]$\lambda \lambda$6717,6731
emission line fluxes for 735 regions $\sim$170 pc in 
diameter. We used the Balmer decrement to calculate nebular dust
extinctions, and correct the observed H$\alpha$ fluxes in
order to measure accurately $\Sigma_{SFR}$ in each region. The THINGS \citep{walter08}
HI 21cm and BIMA-SONG \citep{helfer03} CO J=1-0 maps of NGC5194
were used to measure the HI and H$_2$ gas surface
density for each region. We used a new Monte Carlo method for fitting
the SFL which includes the intrinsic scatter in the
relation as a free parameter, allows the inclusion of non-detections in
both $\Sigma_{gas}$ and $\Sigma_{SFR}$, and is free of the systematics
involved in performing linear correlations over incomplete data in
logarithmic space. After rejecting regions whose nebular spectrum is
affected by the central AGN in NGC5194, we use the [SII]/H$\alpha$ 
ratio to separate spectroscopically the contribution from the diffuse
ionized gas (DIG) in the galaxy, which has a different
temperature and ionization state from those of H II regions in the
disk. The DIG only accounts for 11\% of the total H$\alpha$ luminosity 
integrated over the whole central region,
but on local scales it can account for up to a 100\% of the H$\alpha$
emission, especially in the inter-arm regions. After removing the DIG
contribution from the H$\alpha$ fluxes, we measure a slope
$N=0.85\pm0.05$, and an intrinsic scatter $\epsilon=0.43\pm0.02$ dex
for the total gas SFL, as shown in Figure \ref{fig-3}. We also measure
a typical depletion timescale $\tau
=\Sigma_{HI+H_2}/\Sigma_{SFR} \approx 2$ Gyr, in good agreement with
recent measurements by \cite{bigiel08}. The atomic gas density
shows no correlation with the SFR, and the total gas SFL in the
sampled density range closely follows the molecular gas SFL.
Integral field spectroscopy allows a much cleaner
measurement of H$\alpha$ emission line fluxes than narrow-band
imaging, since it is free of the systematics introduced by continuum
subtraction, underlying photospheric absorption, and
contamination by the [NII] doublet.

Figure \ref{fig-3} also shows that the disagreement with the previous measurement of a
super-linear ($N=1.56$) SFL in NGC5194 by \cite{kennicutt07} is due to
differences in the fitting method, given the good agreement seen in
the data. Applying our Monte Carlo fitting method to the
\cite{kennicutt07} data yields a slope of $N=1.03\pm0.08$. 
Our results support the recent evidence for a low, 
and close to constant, star formation efficiency (SFE=$\tau^{-1}$) in
the molecular component of the ISM of normal spirals. The data shows an excellent agreement
with the recently proposed model of the SFL by \cite{krumholz09}.
The large intrinsic scatter observed may imply the existence of other
parameters, beyond the availability of gas, which are important at
setting the SFR. 

\section{Conclusions}

VENGA will provide SFR maps from H$\alpha$ emission, which together with
far-UV and 24$\mu$ imaging, and PAH features in the IRS mid-IR
spectroscopy will be used to study systematics in different SFR
estimators, revise their calibrations, and further understand
dust reprocessed radiation. Measuring the star-formation
efficiency (SFE=$\Sigma_{SFR}/\Sigma_{gas}$) throughout disks, and
looking for correlations between SFE and properties like metallicity,
orbital timescale, stellar and gaseous velocity dispersion, stellar
mass density, and local ionization field, will test different
theoretical models for star formation laws and thresholds, and GMC
formation.

Also, VENGA produces metallicity, velocity and velocity dispersion maps for
gas and stars from emission and stellar absorption line fits. It also
yields stellar population (SP) mapping of bulges, bars and disks. SP
fitting benefits from the UV and IR imaging to break
age-metallicity-reddening degeneracies in the SED modeling. Modeling
of the SP will yield the SFR histories constraining the stellar
buildup of galaxies. Metallicity and abundance gradients can constrain
the galaxy merger history. Alpha/Fe ratio gradients test how effective
secular processes are at driving gas inflow and inducing star
formation. If bulge assembly proceeds through reassembly of disk
material, the alpha/Fe enhancement in the bulge should not differ much
from that in the disk. Gas kinematics are being used to
quantify non-circular motions along bars, such as azimuthal or radial
streaming (Figure 2), to test the role that bar-driven inflow plays in
building pseudo-bulges. We will also measure $v/\sigma$ in the
central regions to look for evidence of rotationally supported
pseudo-bulges.

For edge-on galaxies, VENGA will vertically map the extra-planar
diffuse ionized gas (EDIG). The origin of the EDIG is not yet
clear. It may well come from internal feedback phenomena described by
galactic fountain models, as well as it could be accreted from the
surrounding IGM. The kinematics and chemical abundances of EDIG
will be used to unveil its origin, which will translate to a better
understanding of feedback and accretion processes driving galaxy
evolution.

In summary, over the next few seasons VENGA will build an
unprecedented spectroscopic dataset on nearby spiral galaxies. This
dataset, in combination with publicly available data at other
wavelengths, will allow us to conduct a large number of studies
regarding the principal processes involved in galaxy formation and
evolution. The VENGA dataset will also become a valuable public
spectroscopic resource on nearby galaxies available to the astronomical community.

\acknowledgements We would like to thank Director David Lambert and the staff at McDonald Observatory for their financial and techincal support of the VENGA observations. G.A.B. thanks the financial support from Sigma Xi, The Scientific Research Society. The construction of VIRUS-P was possible thanks to the generous support of the Cynthia \& George Mitchell Foundation. N.J.E. and A.H. were supported in part by NSF Grant AST-0607193. This research has made use of the NASA/IPAC Extragalactic Database (NED) which is operated by the Jet Propulsion Laboratory, California Institute of Technology, under contract with the National Aeronautics and Space Administration, and of NASA's Astrophysics Data System Bibliographic Services.

\end{document}